\pdfoutput=1
\documentclass{article}

\usepackage[preprint]{tackling_climate_workshop_style}

\usepackage[utf8]{inputenc} 
\usepackage[T1]{fontenc}    
\usepackage{hyperref}       
\usepackage{url}            
\usepackage{booktabs}       
\usepackage{amsfonts}       
\usepackage{nicefrac}       
\usepackage{microtype}      
\usepackage{graphicx}
\usepackage{multirow}
\usepackage{wrapfig}
\usepackage{caption}
\usepackage{subcaption}
\usepackage{authblk}
\graphicspath{ {./Figures/} }

\title{IceCloudNet: Cirrus and mixed-phase cloud prediction from SEVIRI input learned from sparse supervision}

\author[1,2,*]{\textbf{Kai Jeggle}}
\author[2]{\textbf{Mikolaj Czerkawski}}
\author[3,2]{\textbf{Federico Serva}}
\author[2]{\authorcr \textbf{Bertrand Le Saux}}
\author[1]{\textbf{David Neubauer}}
\author[1]{\textbf{Ulrike Lohmann}}
\affil[1]{Institute for Atmospheric and Climate Science, ETH Zurich, Zurich, Switzerland}
\affil[2]{$\Phi$-lab, European Space Agency (ESA), Frascati, Italy}
\affil[3]{Consiglio Nazionale delle Ricerche - Istituto di Scienze Marine (CNR-ISMAR), Rome, Italy}
\affil[*]{Corresponding author: Kai Jeggle, kai.jeggle@env.ethz.ch}

\begin{document}

\maketitle

\begin{abstract}

Clouds containing ice particles play a crucial role in the climate system. 
Yet they remain a source of great uncertainty in climate models and future climate projections. In this work, we create a new observational constraint of regime-dependent ice microphysical properties at the spatio-temporal coverage of geostationary satellite instruments and the quality of active satellite retrievals. We achieve this by training a convolutional neural network on three years of SEVIRI and DARDAR data sets. This work will enable novel research to improve ice cloud process understanding and hence, reduce uncertainties in a changing climate and help assess geoengineering methods for cirrus clouds.
\end{abstract}

\section{Introduction}

\paragraph{Clouds containing ice} particles cover 22 \% of Earth's surface at any moment \citep{heymsfield_cirrus_2017}. They modulate incoming and outgoing radiation \citep{liou_influence_1986} and contribute to the majority of global precipitation \citep{mulmenstadt_frequency_2015}. Yet, we lack understanding about cloud formation and evolution which leads to large uncertainties in climate projections. Clouds containing ice particles can be distinguished into cirrus clouds and mixed-phase clouds. Cirrus clouds contain only ice crystals, are typically optically thin, occur at high altitudes at temperatures below -38°C \citep{sassen_global_2008}, and have on average a warming effect \citep{heymsfield_cirrus_2017}. Mixed-phase clouds contain a mix of ice crystals and supercooled cloud droplets. They are thicker and usually exert a cooling effect \citep{korolev_mixed-phase_2017}. Key to process understanding are properties such as ice water content (IWC) and ice water path (IWP). The former is a measure of the ice mass per unit volume and is vertically resolved while the latter is its integration along a vertical column. Ice properties are projected to change in a warming climate and may amplify or dampen global warming \citep{lohmann_importance_2018}.

\paragraph{Current observational constraints} are mainly two types of satellite retrievals. On the one hand, multiple studies e.g. \citep{kramer_microphysics_2016, sourdeval_satellite-based_2020,jeggle_understanding_2023} have used polar-orbiting active satellite instruments like CALIPSO's lidar \citep{winker_caliop_2009} and CloudSat's radar \citep{stephens_cloudsat_2002} to analyze microphysical properties of ice clouds. These instruments retrieve cloud properties by emitting electromagnetic waves and measuring the backscattered signal. The main benefit is that active instruments are able to provide a vertical profile of cloud structures. Due to the narrow swath of the satellite overpass CALIPSO and CloudSat have a revisiting time of 16 days which is a much coarser temporal scale than necessary for studying clouds where processes act between seconds to hours \citep{lu_which_2018}. In this work, we will use the DARDAR \citep{cazenave_evolution_2019} data set which combines CALIPSO and CloudSat retrievals. On the other hand, passive geostationary satellite instruments such as SEVIRI onboard the Meteosat satellite \citep{aminou_msgs_2002} retrieve a top-down 2D view of Earth's surface every 15 minutes by measuring the reflected solar radiation and intensities of terrestrial infrared radiation. Previous studies have combined geostationary and actively sensed retrievals for IWP predictions \citep{amell_ice_2022, kox_retrieval_2014}, but the context of cloud regime was not considered unlike in this work.
    
\paragraph{The objective} here is to provide regime-dependent IWP with high spatio-temporal coverage. To this end, we train a convolutional neural network (CNN) that predicts IWP for cirrus and mixed-phase clouds from SEVIRI data which is supervised by co-located IWP retrievals from DARDAR. Unlike previous work, the model proposed here provides insight into both ice regimes independently.

\section{Dataset}

\begin{wrapfigure}{r}{0.3\textwidth}
  \begin{center}
    \includegraphics[width=0.3\textwidth]{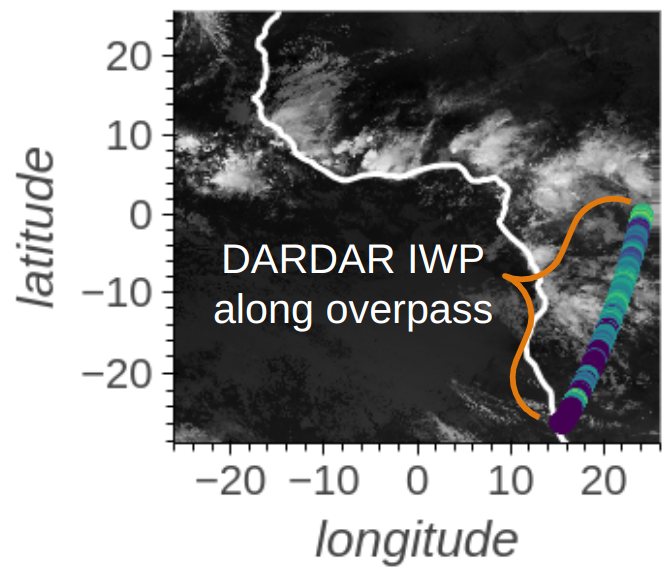}
    \caption{10\textmu m channel for single SEVIRI observation from 2007-10-24 00:12:43 with co-located DARDAR IWP. Note that the DARDAR swath is magnified for better visibility.}
    \label{fig:data_viz}
  \end{center}
\end{wrapfigure}

The dataset used for this study contains three years of multi-spectral images from SEVIRI and vertically-resolved DARDAR IWC swaths. We identify the matching DARDAR overpasses for every SEVIRI image and resample them to the native SEVIRI horizontal resolution of 3km x 3km. Due to the high temporal resolution of 15 minutes and the long revisiting times of DARDAR, any SEVIRI image contains at most one DARDAR overpass, resulting in a minor fraction of a given SEVIRI image that can be co-located with DARDAR data. Note, that many SEVIRI images do not contain any DARDAR overpass. Figure~\ref{fig:data_viz} visualizes an exemplary SEVIRI image with its matching DARDAR swath.  
Instead of using the whole vertical column of DARDAR IWC as target data, we integrate the IWC for cirrus and mixed-phase regimes resulting in an ice-regime dependent IWP. With this approach, we keep a key aspect of vertical cloud structure and ice distribution through the atmosphere while reducing the output from 419 vertical levels to 2. We chose a domain from 30°W to 30°E and 30°N to 30°N, resulting in 1984$\times$1792 pixels per SEVIRI image in its native resolution. For training and validating the neural network, we create non-overlapping patches of size 64$\times$64 and select only the patches that contain co-located DARDAR data, resulting in 160,137 patches for the years 2007-2009. Formally, our dataset is described as:

\begin{itemize}
    \item $X \in \mathbb{R}^{160137 \times 64 \times 64 \times 9}$ for the nine infrared channels of SEVIRI. Three visible channels are omitted as they are only available at daytime.
    \item $Y_{\textrm{sparse}} \in \mathbb{R}^{160137 \times 64 \times 64 \times 2}$ for the IWP of cirrus and mixed-phase regime derived from DARDAR ice water content. Note, that while being on a 64$\times$64 grid, only a small subset of the grid points contain data, i.e. along the DARDAR overpass.
    \item $M = \mathbb{Z}_{2}^{160137 \times 64 \times 64 \times 1}$ for the binary overpass mask indicating the location of the satellite overpass.
\end{itemize}

A considerable challenge from the machine learning perspective is the sparsity of the target data. Yet, our model, once trained and validated, is able to produce IWP predictions for cirrus and mixed-phase clouds at high spatio-temporal coverage for all available SEVIRI images, resulting in full image predictions for ~9.1 million patches of size 64$\times$64.

\section{Methodology}

\subsection{Problem setting}
The outlined task can be formally described as learning a mapping $f: \mathcal{X} = \mathbb{R}^{64 \times 64 \times 9} \rightarrow \mathcal{Y} = \mathbb{R}^{64 \times 64 \times 2}$, where $x \in \mathcal{X}$ is representing SEVIRI channels and $y \in \mathcal{Y}$ the IWP of cirrus and mixed-phase clouds. Note that at training time, a prediction $\hat{y} \in \mathcal{Y}$ is masked to a narrow swath using the corresponding overpass mask $m \in M$ and the loss $\mathcal{L}(\hat{y}_{\textrm{sparse}},y_{\textrm{sparse}})$ is calculated on the sparse data only. 

\subsection{IceCloudNet architecture}
\begin{figure}[ht!]
  \centering
  \includegraphics[width=0.75\textwidth]{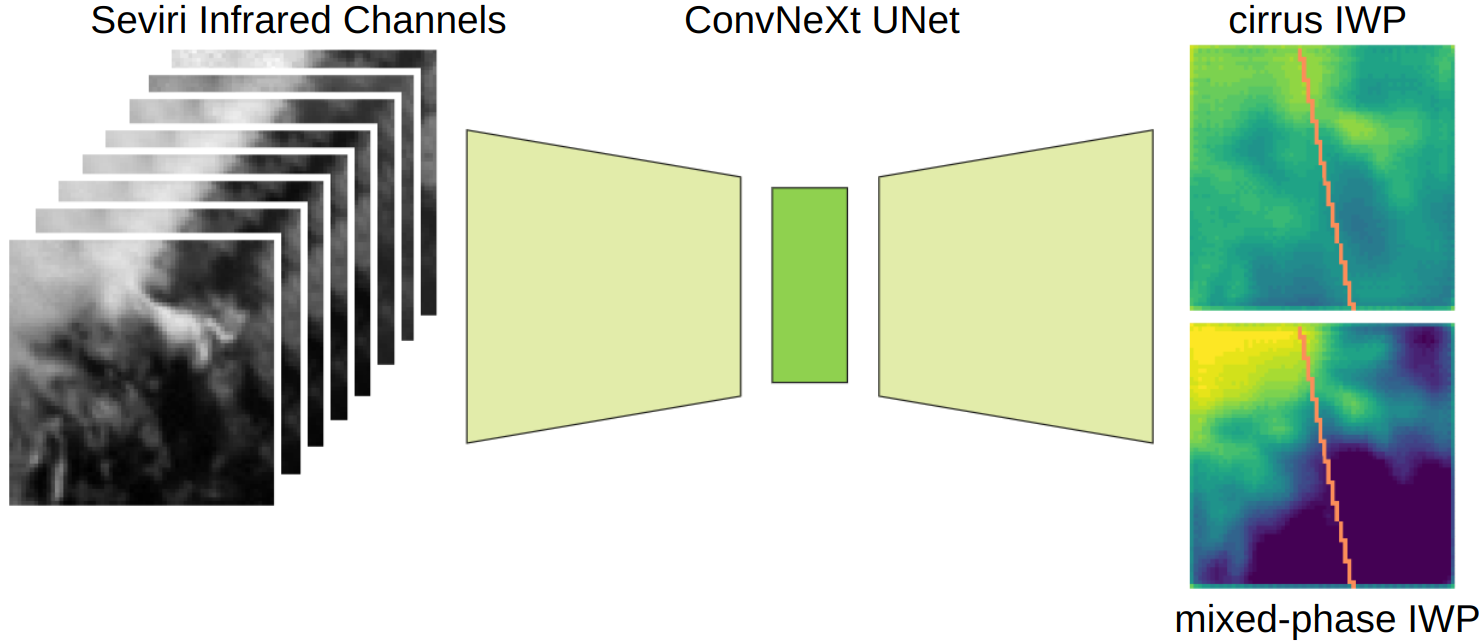}
  \caption{IceCloudNet architecture with sample inputs and predictions. At training, ground truth target data are narrow DARDAR IWP swaths shown as orange lines which are the only supervision for the network. At inference, dense images of the size of the SEVIRI inputs are predicted.}
  \label{fig:architecture}
\end{figure}

The backbone of IceCloudNet is a U-Net \citep{ronneberger_u-net_2015} architecture made up of ConvNeXt \citep{liu_convnet_2022} blocks. ConvNeXt blocks are state-of-the-art convolutional modules based on the ResNet~\citep{he_deep_2015} architecture, improved with multiple macro and micro design choices inspired by transformer models, such as ViT \citep{dosovitskiy_image_2021}. Figure \ref{fig:architecture} visualizes the processing pipeline, where SEVIRI input data $x \in X$ is fed into the network predicting the values of cirrus and mixed-phase IWP $\hat{y} \in \mathcal{Y}$. The orange lines in the predictions show the DARDAR overpass mask $m \in M$, i.e. where target data for supervision is available. Despite being trained only on sparse target data, the model learns to predict the full spatial image, expanding the spatio-temporal coverage of cirrus and mixed-phase IWP considerably.


\subsection{Experimental setup}

We split the dataset into train, validation, and test splits in a 70\%, 20\%, 10\% proportion. In order to prevent spatio-temporal auto-correlation in the splits, data from the same day will be assigned to only one split. We transform $Y_{\textrm{sparse}}$ logarithmically with base 10 and normalize $X$ with the training set statistics. We randomly apply rotations of 90° multiples for data augmentation and train IceCloudNet for 100 epochs and batch size 32 using the Adam optimizer and learning rate of 10\textsuperscript{-6}.

\section{Results}

We compare the predictive performance of IceCloudNet with two baseline models: linear regression and gradient boosted regression trees (XGBoost) \citep{chen_xgboost_2016}. Unlike CNNs, these models are not capitalizing on the spatial structure of the input data, but are predicting the cloud properties pixel by pixel. 

\begin{table}[ht!]
\small
\caption{IceCloudNet performance on independent test set. Note that MAE and correlation are computed only on available DARDAR ground-truth and are calculated on log transformed target variable.}
\label{table:results}
\centering
\begin{tabular}{@{}llcccc@{}}
\toprule
& \textbf{Cloud regime}  & \textbf{MAE} $\downarrow$  & \textbf{Pearson Correlation} $\uparrow$&\textbf{ Accuracy (\%)} $\uparrow$ &  \\ \midrule
\multirow{2}{*}{\textbf{Linear regression}} & cirrus      & 0.83 & 0.75 & 77     &  \\
                                            & mixed-phase & 0.66 & 0.75 & 84     &  \\ \midrule
\multirow{2}{*}{\textbf{XGBoost}}           & cirrus      & 0.74 & 0.75 & 77     &  \\
                                            & mixed-phase & 0.60 & 0.78 & 84     &  \\ \midrule
\multirow{2}{*}{\textbf{IceCloudNet}}       & cirrus      & 0.49 & 0.82                & 86     &  \\
                                            & mixed-phase & 0.47 & 0.83                & 88     &  \\ \bottomrule
\end{tabular}
\end{table}

\begin{figure}[ht!]
  \centering
  \includegraphics[width=\textwidth]{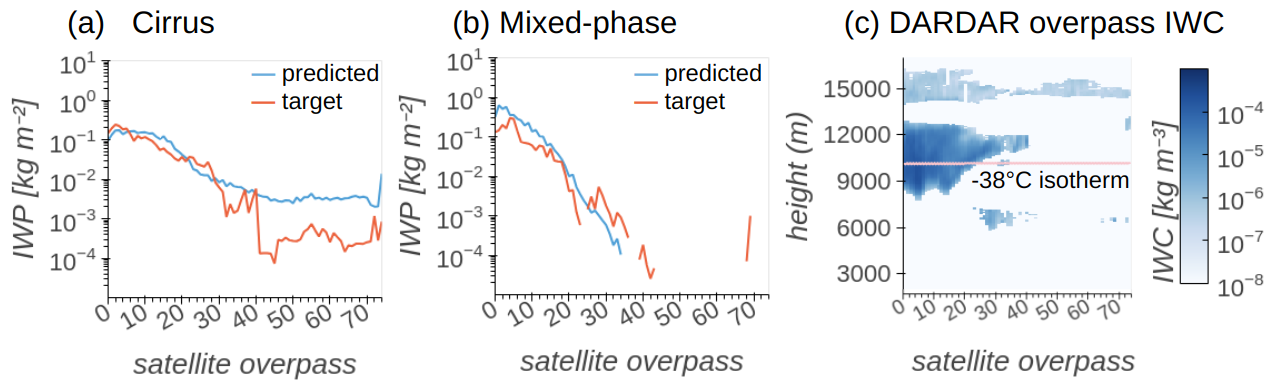}
  \caption{Sample predictions and ground truth along overpass for single patch shown in Figure 2. Panel (c) visualizes vertically resolved IWC ground truth and the -38°C isotherm retrieved from ERA5 \citep{hersbach_era5_2018}. The x-axis is an incremental counter for pixels along the overpass.}
  \label{fig:single_patch}
\end{figure}

Regression and classification metrics can only be calculated where DARDAR ground truth data is available. The performance on the test set is reported in Table \ref{table:results}. The accuracy is calculated based on a post-processed cloud mask where a pixel contains a cloud if the IWP > 10\textsuperscript{-5} kg m\textsuperscript{-2}. IceCloudNet outperforms the baseline models as it is able to learn from the spatial structure in the input data. Figure \ref{fig:single_patch} illustrates the predicted and target variables along the satellite overpass for a single patch (a,b) as well as the ground truth IWC (c) from which the ice regime dependent IWP is derived. All ice above the height of the -38°C isotherm (horizontal line in panel (c)) is in the cirrus regime, everything below in the mixed-phase regime. Despite the complex multi-layer cloud scene, IceCloudNet is able to separate the cirrus and mixed-phase regimes and successfully quantify the IWP for both regimes for the majority of the scene. More samples are shown in Appendix \ref{appendix:prediction_samples}. 

\begin{figure}[ht!]
  \centering
  \includegraphics[width=\textwidth]{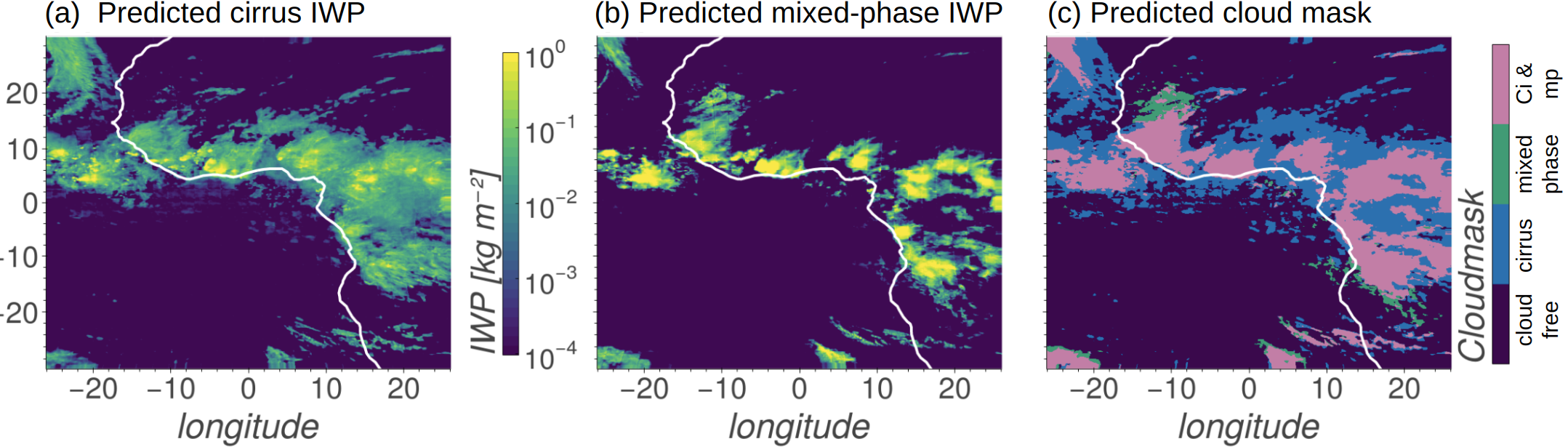}
  \caption{Sample prediction of IceCloudNet for cirrus (a) and mixed phase (b) IWP and cloud mask (c). One channel of the SEVIRI input for this prediction is shown in \ref{fig:data_viz}}
  \label{fig:prediction_sample}
\end{figure}

In Figure \ref{fig:prediction_sample}, we show the prediction of IWP (a-b) of a full SEVIRI observation. Additionally, the post-processed cloud mask is shown in (c), which allows studying cirrus cloud origin \citep{gasparini_cirrus_2018}. Figure~\ref{fig:data_viz} displays SEVIRI input and the available DARDAR ground truth for this observation. With IceCloudNet we are able to expand the information of ice-regime dependent IWP from a narrow swath to the full image. We note that adding geographical information via a topographic map did not improve the prediction performance.

\section{Conclusions}


We introduce a new way to obtain high-quality predictions for microphysical properties of cirrus and mixed-phase clouds with high spatio-temporal coverage. 
Trained on geostationary SEVIRI data and retrievals of actively measured DARDAR data, our machine learning based approach allows to supply the community with a new observational constraint that will enable novel research on ice cloud formation and improve understanding of the microphysical process by tracking and studying cloud properties through time and space, even beyond the lifetime of recently-ended satellite missions underlying DARDAR. We show that IceCloudNet can learn from sparse data and significantly outperforms baseline models. New findings enabled by our work will help to improve climate models reduce climate projection uncertainty and help assess the risk of geoengineering methods.


\begin{ack}
This research was supported by grants from the European Union’s Horizon 2020 research and innovation program iMIRACLI under Marie Skłodowska-Curie grant agreement No 860100 and Swiss National Supercomputing Centre (Centro Svizzero di Calcolo Scientifico, CSCS; project ID s1144). KJ is grateful for the opportunity to being hosted as a visiting researcher at ESA $\Phi$-lab during spring 2023 which led to the initiation of this research project.
\end{ack}

\newpage

\bibliographystyle{plain}
\bibliography{references.bib}

\newpage

\appendix
\section{Prediction samples}\label{appendix:prediction_samples}
\renewcommand{\thefigure}{A\arabic{figure}}
\setcounter{figure}{0}
\begin{figure}[ht!]
     \centering
     \begin{subfigure}[b]{0.9\textwidth}
         \centering
         \includegraphics[width=\textwidth]{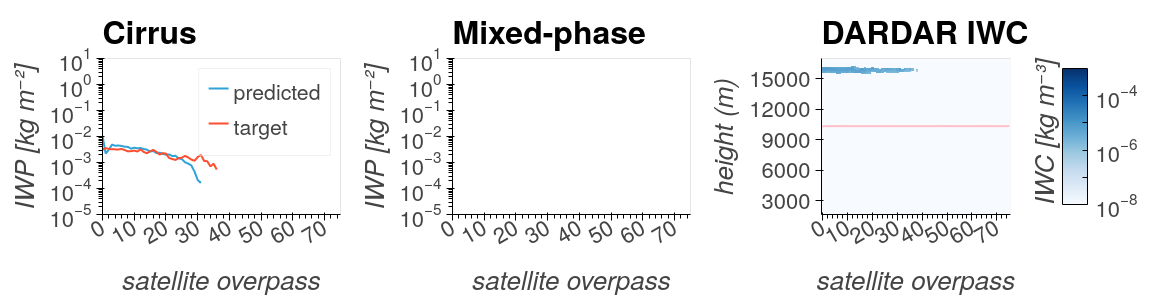}
         \caption{Continuous narrow cirrus cloud}
         \label{fig:a}
     \end{subfigure}
     \hfill
     \begin{subfigure}[b]{0.9\textwidth}
         \centering
         \includegraphics[width=\textwidth]{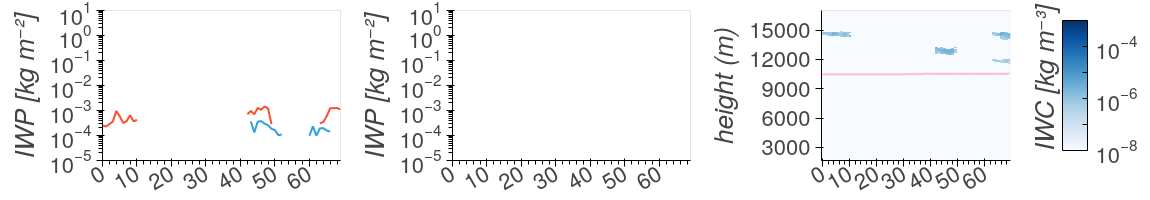}
         \caption{Cirrus clouds with small spatial extent}
         \label{fig:b}
     \end{subfigure}
     \hfill
     \begin{subfigure}[b]{0.9\textwidth}
         \centering
         \includegraphics[width=\textwidth]{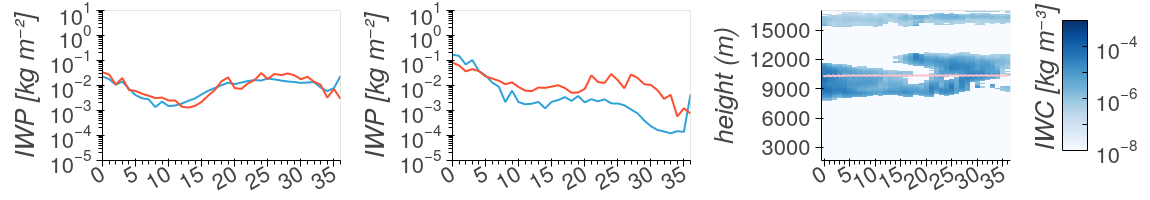}
         \caption{Multi-layer cloud scene}
         \label{fig:c}
     \end{subfigure}
     \hfill
     \begin{subfigure}[b]{0.9\textwidth}
         \centering
         \includegraphics[width=\textwidth]{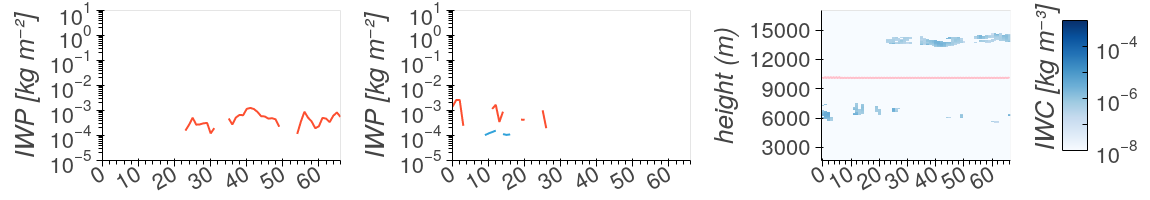}
         \caption{Small blobs of mixed-phase and cirrus clouds}
         \label{fig:d}
     \end{subfigure}
     \hfill
     \begin{subfigure}[b]{0.9\textwidth}
         \centering
         \includegraphics[width=\textwidth]{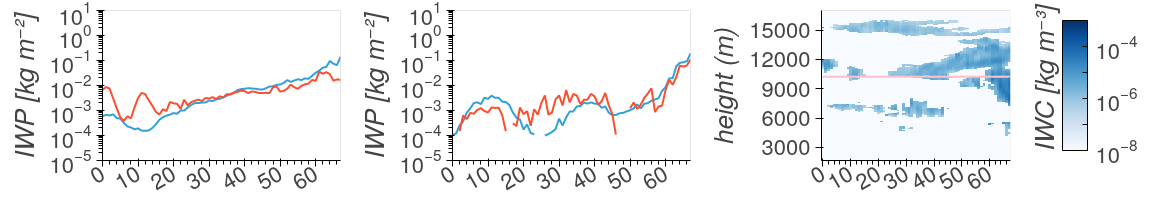}
         \caption{Chaotic cloud scene containing both cirrus and mixed-phase regimes}
         \label{fig:five over x}
     \end{subfigure}
     \hfill
     \begin{subfigure}[b]{0.9\textwidth}
         \centering
         \includegraphics[width=\textwidth]{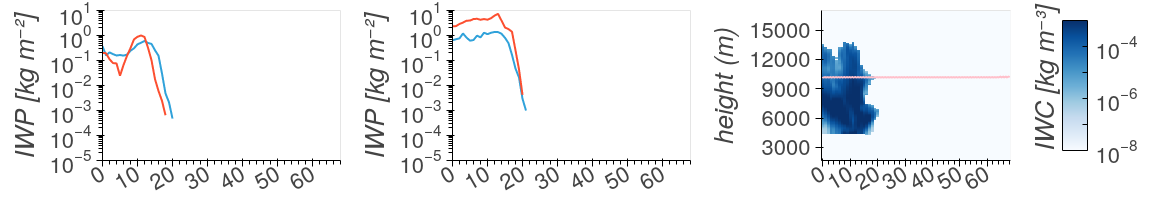}
         \caption{Mixed-phase cloud with large vertical extent and cirrus top}
         \label{fig:five over x}
     \end{subfigure}
        \caption{Sample predictions of IceCloudNet (blue lines) and DARDAR ground truth (red lines) for cirrus (left) and mixed-phase (center) IWP. Each row represents a random sample from the test set. The right plot in each row shows the DARDAR IWC ground truth along the overpass, i.e the vertically resolved variable and is displayed to provide insight into the structure of the cloud scene. The horizontal orange line represents the ERA5 derived -38°C isotherm which acts as a border between cirrus and mixed-phase regimes. The x-axis an incremental counter for pixels along the overpass which can vary depending on the overpass angle.}
        \label{fig:appendix_prediction_samples}
\end{figure}

Figure \ref{fig:appendix_prediction_samples} shows IceCloudNet predictions for randomly sampled patches of the test set. From qualitative inspection, we note that IceCloudNet is able to detect and quantify regime-dependent IWP for larger cloud structures (e.g. a,c,f), even if the scene is very chaotic (e.g. d), but may struggle on clouds with small horizontal extent (e.g. b,d). From a climate impact perspective, small clouds have minor relevance compared to large cloud systems. Nonetheless, we aim to improve the performance of IceCloudNet on smaller clouds in future versions.

\end{document}